\documentclass{sig-alternate}
\usepackage{multirow}
\usepackage{url}
\usepackage{enumitem}

\begin{document}

%\conferenceinfo{SIGIR'16,}{17-21 July, 2016, Pisa, Italy}
\conferenceinfo{Neu-IR '16 SIGIR Workshop on Neural Information Retrieval,}{July 21, 2016, Pisa, Italy}
\global\boilerplate={Permission to make digital or hard copies of part or all of this work for personal or classroom use is granted without fee provided that copies are not made or distributed for profit or commercial advantage and that copies bear this notice and the full citation on the first page. Copyrights for third-party components of this work must be honored. For all other uses, contact the authors.}
\newtoks\copyrightetc
\global\copyrightetc{\copyright~2016 Copyright held by the authors}

\title{Selective Term Proximity Scoring Via BP-ANN}

\numberofauthors{3} %  in this sample file, there are a *total*
\author{
Ju Yang,
Jiancong Tong,
Rebecca J.\ Stones,
Zhaohua Zhang,
Benjun Ye,
Gang Wang,
Xiaoguang Liu \\[0.5em]
College of Computer and Control Engineering, Nankai University, Tianjin, China\\[0.7em]
Email: \{yangju,jctong, rebecca.stones82, zhangzhaohua, yebj, wgzwp, liuxg\}@nbjl.nankai.edu.cn
}
% There's nothing stopping you putting the seventh, eighth, etc.
% author on the opening page (as the 'third row') but we ask,
% for aesthetic reasons that you place these 'additional authors'
% in the \additional authors block, viz.

% Just remember to make sure that the TOTAL number of authors
% is the number that will appear on the first page PLUS the
% number that will appear in the \additionalauthors section.

\maketitle
\begin{abstract}
When two terms occur together in a document, the probability of a close relationship between them and the document itself is greater if they are in nearby positions.  However, ranking functions including term proximity (TP) require larger indexes than traditional document-level indexing, which slows down query processing.  Previous studies also show that this technique is not effective for all types of queries.  Here we propose a document ranking model which decides for which queries it would be beneficial to use a proximity-based ranking, based on a collection of features of the query.  We use a machine learning approach in determining whether utilizing TP will be beneficial. Experiments show that the proposed model returns improved rankings while also reducing the overhead incurred as a result of using TP statistics.

\end{abstract}

\category{H.3.3}{Information Storage and Retrieval}Information Search and Retrieval-Search process
%{H.4}{Information Systems Applications}{Miscellaneous}
%A category including the fourth, optional field follows...
%\category{D.2.8}{Software Engineering}{Metrics}[Experimentation, performance measures]

\keywords{Information retrieval, Term proximity, Query Effectiveness Prediction}%, Learning to Rank,}

\section{Introduction}
Search engine users want relevant documents returned quickly when searching. Searching is often performed using an inverted index \cite{yan2009inverted}, which stores a list of occurring terms, and for each term, documents in which that term occurs are recorded, along with term frequency within each document and all the corresponding position information.

The search process for conjunctive queries goes through two main phases: list intersection and ranking \cite{constantinos2013a}. We first find the documents that contain all the query terms, then rank them according to their relevance to the query; ideally, the most relevant documents are returned.  Traditional methods for assessing if a document is relevant to a query use two kinds of features: (a) term-independent features, e.g., PageRank; and (b) term-dependent features.  Term-dependent features focus on term frequency and inverted document frequency. BM25 \cite{robertson1976relevance} is one of the most widely used; given query $q$, the document $d$ is assigned the score:
\begin{equation}\label{eq:BM25_score}
S_{\mathrm{BM}25}(q,d)=\sum_{\text{term } t \in q} w_t\frac{f_{d,t}(k_1+1)}{f_{d,t}+k_1(1-b+\frac{b|d|}{\mathrm{avg}_d})},
\end{equation}
where $k_1$ and $b$ are predefined constants, $|d|$ is the length of document $d$, $\mathrm{avg}_d$ is the average document length in the collection, $f_{d,t}$ is the frequency of $t$ in document $d$, and $w_t$ is the inverse document frequency (idf) of term $t$.

The order in which terms appear in the document and the distance between their locations are both important ranking criteria. Consider the two-term query ``search engine'' for ranking the following toy example documents:
\begin{itemize}[leftmargin=0.7cm]
  \item[$d_1$:] \textit{$\ldots$ word \textbf{search engine} word word word word \textbf{search engine} word word word word word word $\ldots$}
  \item[$d_2$:] \textit{$\ldots$ \textbf{search} word word \textbf{search} word word \textbf{engine} \textbf{search} word \textbf{engine search} word \textbf{engine} word $\ldots$}
\end{itemize}
In this example, document $d_1$ can be regarded as more relevant, despite $f_{d_2,t}>f_{d_1,t}$ for both terms $t \in \{\text{search},\text{engine}\}$. Consequently, there is active research on methods of integrating term proximity (TP) into the usual ``bag of words'' ranking \cite{yan2010efficient}.

TP score has been demonstrated to have an overall positive effect on search quality \cite{%buttcher2005efficiency,buttcher2006term,song2008viewing,
tao2007exploration}. %[[do we need all of these citations?  is Tao and Zhai (2007) by itself enough?]].
However, there are two caveats: (a) some queries return inferior rankings when utilizing TP score, and (b) incorporating TP score increases computational time. This motivates us to propose a model which selects which queries would likely benefit from incorporating TP score into their ranking.

% However, our experiments show that a small number of queries suffer from TP score while others obtain performance improvement. In this paper, we aim to ensure effective and efficient retrieval by selecting the queries of which the TP score should be taken into consideration when performing the ranking.

The remainder of the paper is organized as follows: Section~\ref{se:related} summarizes related work on proximity ranking models. Section~\ref{se:TP_model} introduces our proposed method. Section~\ref{se:exp} details the experiment setup and presents the results. In Section~\ref{se:conc}, we present our conclusions and future research directions are suggested.

\section{Related Work}\label{se:related}

There are two types of models using proximity in ranking: (a) complex ranking functions that combine hundreds of features (TP being one of them) using sophisticated machine learning techniques \cite{liu2007letor}, and (b) variations of
%that add a new factor \cite{metzler2005markov}
the classic ranking models. While the former achieves more effective results than the latter, it is sometimes too computationally expensive to use. Recent work has explored approaches to achieve a better balance between retrieval effectiveness and efficiency \cite{wang2011cascade}. However, in this paper, we aim to treat each query flexibly, so we focus on the latter.

Rasolofo and Savoy \cite{rasolofo2003term} proposed a TP-based ranking scheme BM25TP, a modified version of BM25 \eqref{eq:BM25_score}, which incorporates term proximity (a similar scheme was presented by B{\"u}ttcher \textit{et al}.\ \cite{buttcher2006term}). In BM25TP, the rank of document $d$ is given by:
\begin{equation}\label{eq:BM25TP_score}
\mathrm{S}^\text{RS}_\text{BM25TP}(q,d)=S_\text{BM25}(q,d)+\mathrm{S}^\text{ACC}_\text{TP}(q,d)
\end{equation}
\begin{equation}\label{eq:ACC_score}
S^{\text{ACC}}_{\text{TP}}(q,d)=\sum_{t\in Q} \min\{1,w_t\} \frac{\mathrm{acc}_d(t) (k_1+1)}{\mathrm{acc}_d(t)+k_1\big(1-b+\frac{b|d|}{\mathrm{avg}_d}\big)},
\end{equation}
where $\mathrm{acc}_d(t)$ denotes the proximity accumulator for term $t$:
\begin{equation*}
\mathrm{acc}_d(t)=\sum_{s \neq t} w_t\,\mathrm{tpi}_d(t,s)
%\mathrm{acc}(t) := \mathrm{acc}(t)+w_t\,\mathrm{tpi}(t,s)
\end{equation*}
and
\begin{equation*}
\mathrm{tpi}_d(t,s)=\sum_{\substack{\text{occurrence } o(t) \text{ of } t \\ \text{in document } d}}\mathrm{dist}(o(t),s)^{-2}
\end{equation*}
for each given term pair $(t,s)$, where $t \neq s$, and $\mathrm{dist}(o(t),s)$ is the number of terms between the position of $o(t)$ and the position of the preceding occurrence of the term $s$.

An assortment of other methods of utilizing TP in ranking have been studied.  Akritidis \textit{et al}.\ \cite{akritidis2012improved} not only takes into account term proximity, but also the order of terms in the query. Zhu \textit{et al}.\ \cite{zhu2009effective} put forward some new ideas based on web page structure and set estimation rules for early termination to speed up top-$k$ computation. %[[this doesn't seem accurate; what did this achieve?]].
Svore \textit{et al}.\ \cite{svore2010good} and Song \textit{et al}.\ \cite{song2008viewing} utilized TP in the form of ``spans'' to improve the accuracy of ranking functions. Tao and Zhai \cite{tao2007exploration} introduced five measures and combined them with an existing retrieval model with two newly designed  heuristic constraints; their experiments showed significant performance improvement on the KL-divergence language model and the BM25 model. Lv and Zhai \cite{lv2009positional} presented four proximity-based density functions to estimate different positional language models (PLMs), namely the Gaussian, triangular, cosine, and circle. Metzler \textit{et al}.\ \cite{bendersky2010learning,metzler2005markov} developed a general Markov random field (MRF) retrieval model that captures various kinds of term dependencies: full independence, sequential dependence, and full dependence. Cummins and O'Riordan \cite{cummins2009learning} outlined an extensive list of possible term proximity measures, and incorporated them to the original framework by machine learning methods.

% Features that are more complicated than square distances and new adding methods have also been presented.

Different queries benefit from different proximity features and methods \cite{cummins2009learning,lu2014effective,svore2010good}. Moreover, a too-complicated ranking formula may be a burden to use, both for the operator and by requiring too much overhead. This motivates us to propose a method where TP statistics are utilized only when they are most useful.

% To the best of our knowledge, this study is the first to propose filtering queries before ranking [[surely someone would have done it before...?]].

\section{Selective TP model}\label{se:TP_model}

\subsection{Features considered}

Table~\ref{ta:features} lists the features considered in this work. These features can be roughly divided into two categories: (a) query dependent, and (b) term dependent. Term dependent features are divided into two subcategories: frequency-based and position-based.

The query dependent features we include is the number of documents related to the query. The inverted document frequency (idf) indicates the overall importance of a term, and we utilize its statistics: mean, min, max, and sum; the sum of squared idfs, and the sum of squared differences between ascendant or descendant idf values between consecutive terms in a query\footnote{Ascendant (resp.\ descendant) idfs refer to consecutive term pairs whose first idf value is less (resp.\ greater) than the second.}. The position-based features include the average position of a term in a document, averaged over all documents, which we call the general position (abbreviated pos). Most pos statistics used are analogous to the idf statistics.  These features vary in their ability to distinguish queries from each other; we specify the features actually used in Section~\ref{se:exp}.
% And we will specify the features truly used in our experiments a bit later.
\begin{table}[t]
\caption{Summary of query features used}
\label{ta:features}
\centering
%\resizebox{\columnwidth}{!}{
\begin{tabular}{|l|}
\hline
\textbf{query features} \\
\hline
number of relevant documents\\
\cline{1-1}
\textbf{term frequency features} \\
\cline{1-1}
mean; min; max; sum idf \\
sum of squared idfs \\
sum of squares of ascendant idfs \\
sum of squares of descendant idfs \\
\cline{1-1}
\textbf{term position features} \\
\cline{1-1}
mean; min; max; sum pos \\
square statistics \\
\cline{1-1}
\end{tabular}
%}
\end{table}

%\begin{table}[t]
%\caption{Summary of query features used}
%\label{ta:features}
%\centering
%\resizebox{\columnwidth}{!}{
%\begin{tabular}{|l|l|}
%\hline
%\textbf{query features} & \textbf{term position features} \\
%\hline
%query length & mean; min; max; sum pos \\
%number of relevant documents & square statistics \\
%\cline{1-1}
%\textbf{term frequency features} & pos squared ascents \\
%\cline{1-1}
%mean; min; max; sum idf & pos span; max-dif \\
%sum of squared idf & var.\ and std.\ dev.\ of pos \\
%\cline{2-2}
%sum of squares of ascendant idf\\
%sum of squares of descendant idf\\
%number of idf ascents \\
%\cline{1-1}
%\end{tabular}
%}
%\end{table}

\subsection{Term proximity score}\label{se:term_prox}

Three of the most popular TP ranking functions, which we test our selective model based on, are introduced here. Two are BM25TP, given by \eqref{eq:BM25TP_score}, and MRF by Metzler \textit{et al}. \cite{metzler2005markov}. MRF is defined by the following ranking function (full details are omitted for space reasons):
\begin{align}
% \mathrm{S_{MRF}} &= \sum_{c\in C(G)}{\lambda}_c f(c)\\
\mathrm{S}_\text{MRF}(q,d) &:= \mathrm{S}_\text{TF}(q,d) + \sum_{c\in O}{\lambda}_Of_O(c) + \sum_{c\in {O \cup U}}{\lambda}_Uf_U(c); \label{eq:MRF_rank} \\
\mathrm{S}_\text{TF}(q,d) &:= \sum_{c\in T}{\lambda}_Tf_T(c). \label{eq:TERM_rank}
\end{align}
where $T$ is the set of 2-cliques involving a query term and a document, $O$ is the set of cliques containing the document node and two or more query terms that appear contiguously within the query, and $U$ is the set for query terms appearing non-contiguously within the query. In this paper, we use an extension model proposed by \cite{bendersky2010learning}.

The third one we use is from Tao and Zhai \cite{tao2007exploration} who instead calculate a term-proximity-based rank by:
\begin{align}
\mathrm{S}^\text{TZ}_\text{EXP}(q,d) &:= \mathrm{S}^\text{TZ}_\text{TP}(q,d)+\mathrm{S_{BM25}}(q,d); \label{eq:TaoZhai_rank}\\
\mathrm{S}^\text{TZ}_\text{TP}(q,d) &:= \log\big(\alpha+\exp(-\mathrm{min\_dist}(q,d))\big) \nonumber
\end{align}
where $\mathrm{min\_dist}(q,d)$ is the minimum distance between any occurrence of any two query terms in document $d$, and $\alpha$ is a parameter.  Tao and Zhai state that \eqref{eq:TaoZhai_rank} provides stable performance when $\alpha$ is set to 0.3, which we use for our experiments.

% Compared with the variants of classical ranking function above, the term dependency model [[what model is this?]] is worse when it comes to parameter estimation.
% A theoretical foundation may be that
Some studies have identified a decaying function between two words to calculate the strength of their association \cite{gao2002resolving,vechtomova2006study,yuret1998discovery}.

% Becky: I got rid of this (it doesn't make sense to me).
% and assume TP score is measured better using a different way from term frequency factor.

\subsection{Our approach}

As our ranking functions, we use \eqref{eq:MRF_rank} and a generalized combination of \eqref{eq:BM25TP_score} and \eqref{eq:TaoZhai_rank}:
\begin{align}
\mathrm{S_{EXP}}(q,d) &:= \epsilon\,\mathrm{S^{\text{TZ}}_{TP}}(q,d)+(1-\epsilon)\,\mathrm{S_{BM25}}(q,d); \nonumber
\\
\mathrm{S_{BM25TP}}(q,d) &:= \beta\,\mathrm{S^{ACC}_{TP}}(q,d)+(1-\beta)\,\mathrm{S_{BM25}}(q,d).\label{eq:our_rank}
\end{align}
Parameters $\epsilon$ and $\beta$ are used to adjust the weighting of BM25 and the TP score. We test $\epsilon,\beta \in \{0.1,0.2,\ldots,0.9\}$ for the two query sets in our experiments, and choose the parameters leading to the best mean average precision (MAP).

The MAP of each query is used to evaluate the performance of the two ranking models. If a query gets better results using \eqref{eq:our_rank} than \eqref{eq:BM25_score} or \eqref{eq:MRF_rank} than \eqref{eq:TERM_rank}, its features will be labeled as 1, otherwise 0. These results are used to train a (supervised) classifier, determining whether or not using TP score is likely to benefit the document rankings for arbitrary queries.

We use a Back Propagation Artificial Neural Network (BP-ANN) to build our selective TP model, because of its powerful learning ability and rapid forecasting speed.
%\begin{itemize}[leftmargin=0.5cm]
%\item\textbf{BP-ANN}:
BP-ANN \cite{rumelhart1988learning} %\cite{linl996neural} % \cite{russell1995modern} [[is this a good reference for BP-ANN?]]
%are simplification of what is currently known about the physical structure and its operation.
uses a back-propagation algorithm to modify the internal network weights during the training process. In our experiment, we establish a one-node (denoting the query type) output layer BP-ANN, which contains one hidden layer and whose input nodes are query features.%  , hidden layer and output layer) for influence of TP score on queries.
%\item\textbf{GBDT}:
%Gradient Boost Decision Tree (GBDT) \cite{friedman2001greedy}, %of which the key is to choose appropriate attributes [[I don't understand]],
%is a useful method of generating effective classification models.
%%is a widely used cluster analysis method. As a perfect quantitative method,
%It constructs iteratively an ensemble of weak decision trees through boosting. %is adaptable, practicable and can produce highly accurate models.
%We use a python GBDT implementation from scikit-learn\footnote{\url{http://scikit-learn.org/}} in our experiment.
%\end{itemize}

\section{Experiments}\label{se:exp}

All experiments are performed on the GOV2 data set using Porter stemming. We use the query set MQ2007 and MQ2008 for evaluation. The BM25 scores used in this paper are extracted from LETOR4.0\footnote{\url{http://research.microsoft.com/en-us/um/beijing/projects/letor/}}.

\begin{figure}[htp]
\centering
\includegraphics[width=5.5cm]{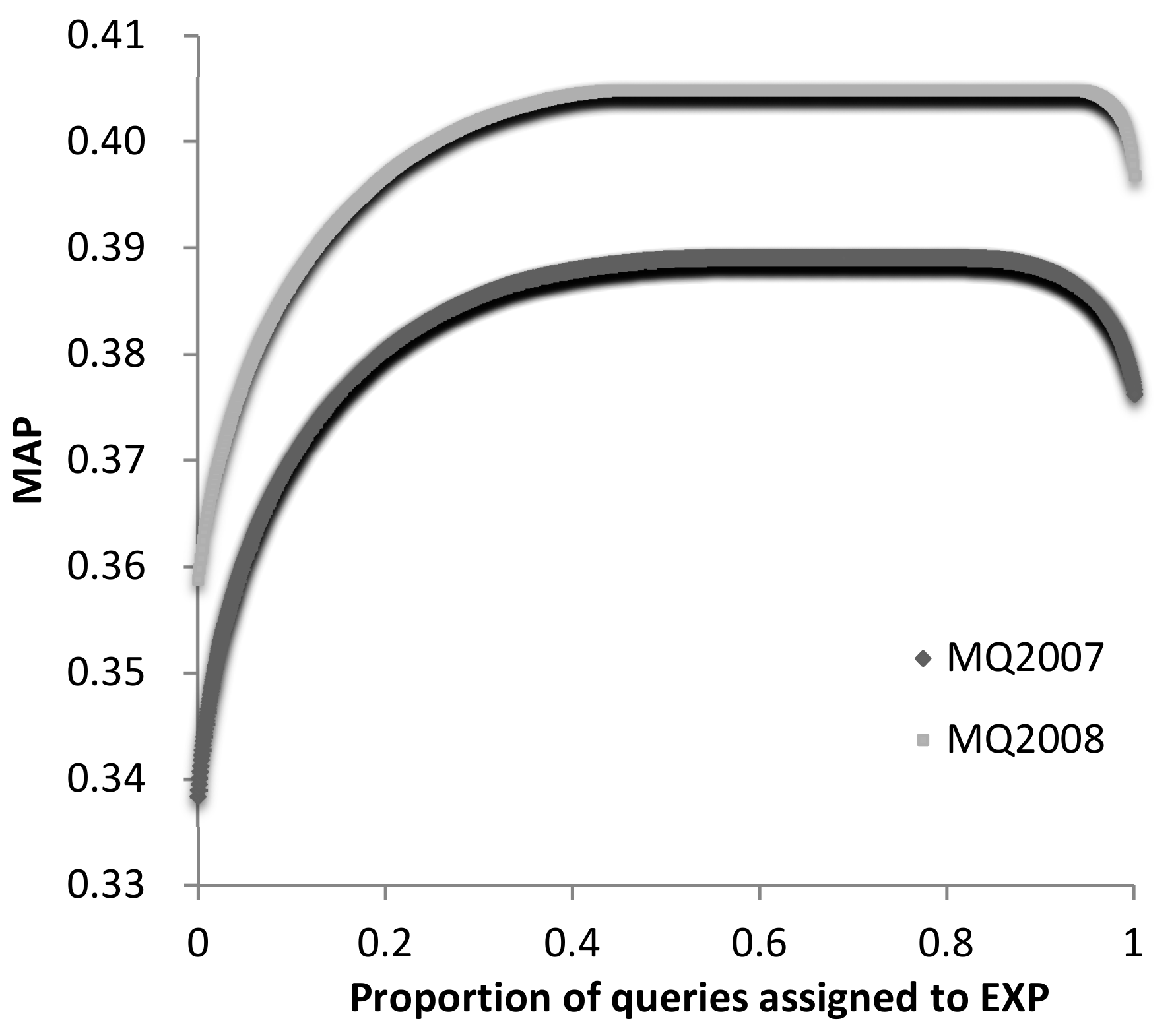}
\includegraphics[width=5.5cm]{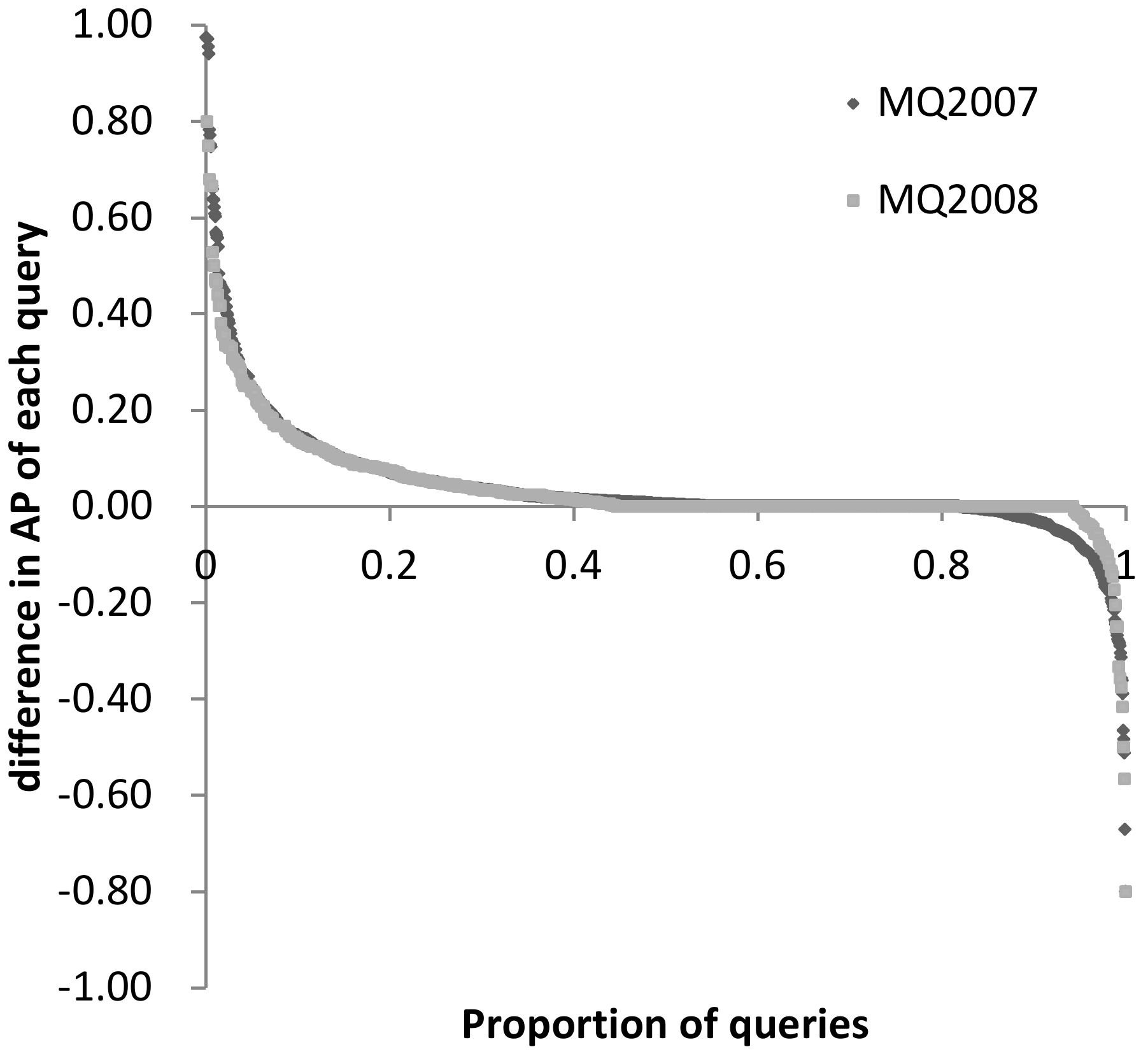}
\caption{MAP on the two query sets as the proportion of queries using TP score varies; Sorted difference in average precision of each query}
%\includegraphics[width=6.5cm]{ndcg_excel}
%\caption{MAP and Mean NDCG on two query set, respectively.}
\label{fig:figuremap}
\end{figure}

Figure~\ref{fig:figuremap} shows the MAP values of the rankings for queries, as the proportion of queries using TP (EXP-score) varies. The queries are sorted by how beneficial it would be to use TP scores in their ranking, with the most benefited coming first. Figure~\ref{fig:figuremap} shows that using TP scores does not always improve retrieval quality (assigning more than around 40\% has no benefit). We also test methods assigning a label $0$ or $1$ to a query randomly, which again shows that naively increasing of proportion of queries utilizing TP will not necessarily result in a performance improvement.

Only relevant features are necessary in the BP-ANN model construction, so we remove unnecessary features. To determine feature importance, we combine statistical methods (ranksum, z-score, and $\chi$-squared), a searching algorithm (decision tree), and a feature weight algorithm (relief).

We find that max idf, sum idf, and sum of squared idfs have relatively more importance among the term frequency features, and max pos, min pos, sum pos, and mean pos have relatively more importance among the term position features. As such, we primarily use max pos, min pos, sum pos, and mean pos for EXP; sum idf, max idf, and min pos for MRF; and min idf, sum of squared idfs, sum of squared differences between descendant idfs, and sum pos for BM25TP.

As other researchers have observed \cite{cummins2009learning,huston2014comparison,lu2014effective,svore2010good}, we also find that query length is an important factor in distinguishing whether or not using TP score will be beneficial, so we train independent models for different query length. Because of their effective performance in a more systematic study \cite{azzopardi2009query}, we consider queries with 3 to 5 terms.

\begin{figure}[htp]
\centering
\includegraphics[width=7.5cm]{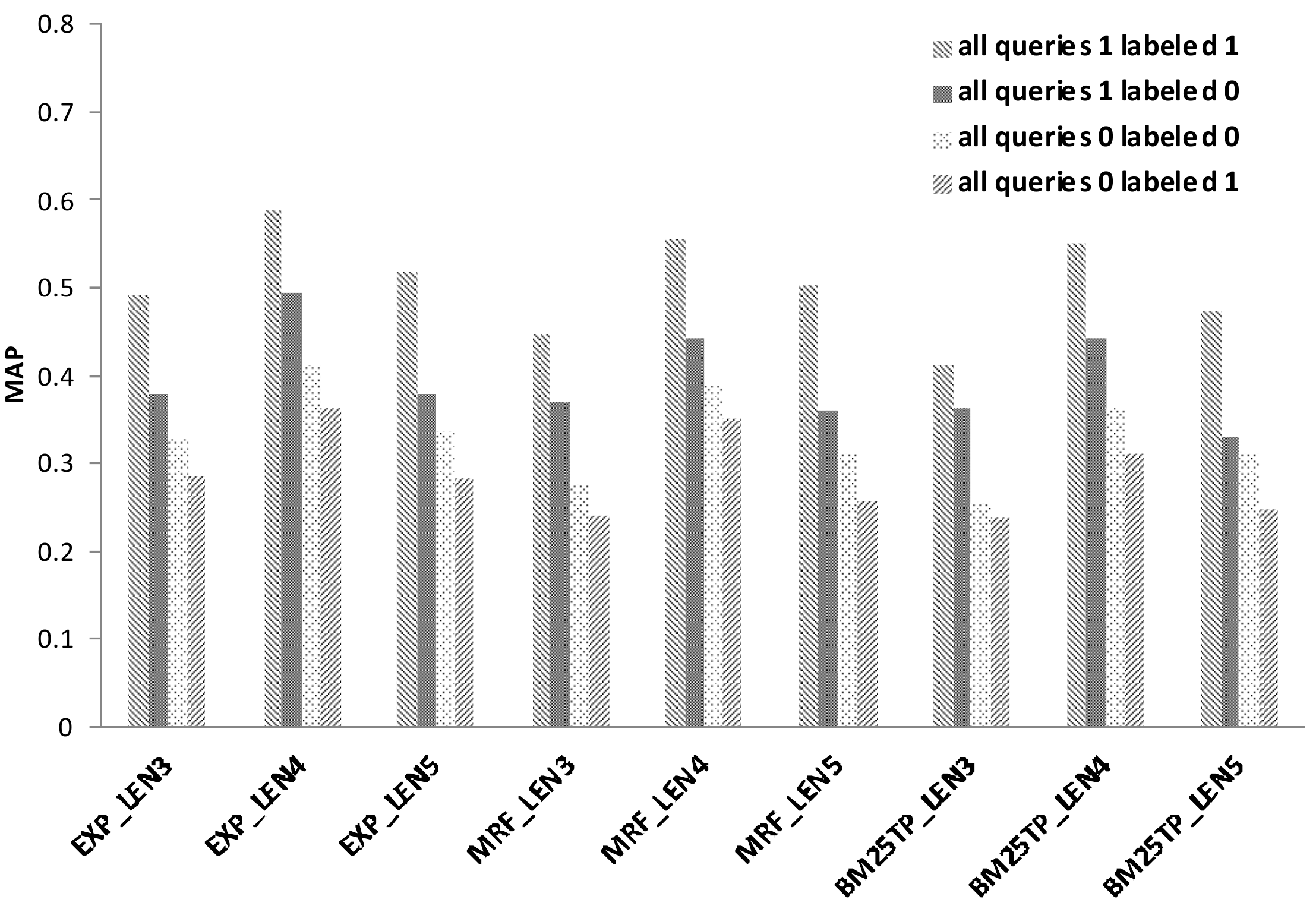}
\caption{MAP degradation caused by wrong judgement for different query sets}
\label{fig:figuremapdeg}
\end{figure}

After each query is labeled, we train a neural network on 70\% of all the data and test the effectiveness and efficiency of the model on the remaining data.  In our BP networks, the input layer has 3 to 5 features and the output layer has 1 node. The maximum number of iterations is set to 1000 and the learning rate of the network is 0.01. We choose the sigmoid function
\[f(x) = \frac{1}{1 + e^{-x}}\]
as the activation function and test the performance of different hidden layer nodes. All the networks aim to correctly predict queries labeled 1 as much as possible, motivated by Figure~\ref{fig:figuremapdeg} which indicates that mispredicting queries labeled 1 is consistently worse than other mispredictions.  The bias on mispredictions is used as a reference for process parameter adjustment in the network. The number of hidden layer nodes used and its momentum coefficient $\alpha$ are listed in Table~\ref{ta:netParameter}, along with the precision and recall values on the training and test data.

%contains 48 nodes for EXP\_len3, 46 nodes for EXP\_len4, 40 nodes for MRF\_len3, 26 nodes for BM25TP\_len3 and 47 nodes for BM25TP\_len4.
%[[The experiment is repeated five times to ensure the results are reliable and each time we use a training set chosen on a random basis (for space reasons, we present the results of one of these five results)]]. %Table~\ref{ta:query num} lists the number of queries in the query set, and the number of queries which would or would not benefit from TP score.

We compare three TP-based rankings for each TP score: \texttt{\_tpAll} (where TP is always used for ranking), \texttt{\_tpS} which calculates TP score depending on BP-ANN predictions (the proposed ranking method), and \texttt{\_oracle}, a theoretically perfect situation where we know \textit{a priori} whether or not a query would benefit from TP score.  For comparison, we also include a non-TP-based ranking \texttt{\_tpNo} given by \eqref{eq:BM25_score} and \eqref{eq:TERM_rank}.

\begin{table*}[htp]
\centering
\caption{Parameters of BP-ANN and the precision and recall values on the training and test data.}
\label{ta:netParameter}
\begin{tabular}{|c|c|c|c|c|c|c|c|c|c|}
\hline
% \textbf{TP Score}
& \multicolumn{3}{|c|}{\texttt{EXP}}& \multicolumn{3}{|c|}{\texttt{MRF}}& \multicolumn{3}{|c|}{\texttt{BM25TP}}\\
\hline
\hline
Query len.&3&4&5&3&4&5&3&4&5 \\
\hline
\hline
Prec.\ on train&59.29\%&61.74\%&61.15\%&53.48\%&60.26\%&56.96\%&54.02\%&49.77\%&42.18\%\\
\hline
Recall on train&89.73\%&99.39\%&94.12\%&91.25\%&90.97\%&91.84\%&90.38\%&87.60\%&91.18\% \\
\hline
Prec.\ on test&69.30\%&72.53\%&64.62\%&53.98\%&62.89\%&55.88\%&51.40\%&51.06\%&47.37\% \\
\hline
Recall on test&98.80\%&95.65\%&95.45\%&88.41\%&92.42\%&95.00\%&82.09\%&92.31\%&90.00\% \\
\hline
\hline
\#hidden nodes&43&58&47&45&39&47&52&10&20\\
\hline
$\alpha$&1&0.85&0.85&0.35&0.9&0.85&0.95&0.25&0.75 \\
\hline
\end{tabular}
\end{table*}

%Experiments are performed on a single machine with Intel Core(TM) i7 CPU @ 3.40GHz, 32GB RAM, and 1.1TB local disk. The operating system is CentOS 6.4.

% The evaluation consists of two parts: the first part is to test the correctness of query classification; the second part is to measure the quality of the results and compare response times of different score functions.

% \subsection{Results}

We use three methods for measuring the quality of the rankings: MAP, precision for top-$k$ results, and Mean Normalized Discounted Cumulative Gain (Mean NDCG), given in Table~\ref{ta:GOV_res}.  We also list the number of queries benefiting from calculation of TP score and the throughput.

% For the various ranking methods on the GOV2 data set for the query set MQ2007, Table~\ref{ta:GOV_res} presents (a) the MAP, P@1, P@3, P@10 and Mean NDCG results and (b) the number of queries benefiting from calculation of TP score.%throughput.

Table~\ref{ta:GOV_res} shows that the TP rankings (\texttt{\_tpAll} and \texttt{\_tpS}) consistently exhibit significantly better rankings than without TP (\texttt{\_tpNo}).  We also see that the selective model (\texttt{\_tpS}) returns slightly better rankings than \texttt{\_tpAll} while having better throughput.  In terms of MAP, we see that MRF is consistently superior to the other ranking formulas.  However, \texttt{\_tpS} used with EXP is the nearest to the corresponding \texttt{\_oracle} (the best possible MAP). Further, we can also find that \texttt{\_tpS} shows a better performance (vs.\ \texttt{\_tpAll}) in terms of $k=1$ precision, which is a critical measure for exact queries, such as queries restricted to web sites.

% becky: said the important part of this before
% Interestingly, regardless of which type of TP score we incorporate, the value of $k=1$ precision is relatively easy to change better and the degree of performance improvement is usually great.

%, and that we gets better performance for BM25TP-Score as query length increase.

\begin{table*}[htp]
\centering
\caption{Performance under the various ranking methods in terms of query length (Q len.), MAP, mean NDCG, precision, and throughput.  We also include the number of queries which uses TP ranking (\#TP-Q).}\label{ta:GOV_res}
\begin{tabular}{|l|c|c|c|c|c|c|c|c|c|}
\hline
& \multirow{2}{*}{Q len.} &\multirow{2}{*}{MAP}&\multirow{2}{*}{Mean NDCG}&{Prec.}&{Prec.}&{Prec.}&\multirow{2}{*}{\#TP-Q}&{Throughput}\\
& & & &{$k=1$}&{$k=3$}&{$k=10$}& & (Q/s) \\

\hline

\multirow{12}{2cm}{\texttt{EXP\_BM25\_tpAll} \texttt{EXP\_BM25\_tpS} \texttt{EXP\_oracle} \texttt{EXP\_tpNo}}

&\multirow{4}{*}{3}&0.4358&0.4497&0.3427&0.3683&0.3615&143&334.56\\
&&{0.4400}&{0.4582}&{0.3776}&{0.3893}&{0.3622}&107&379.71\\
&&0.4481&0.4733&0.3986&0.4033&0.3748&80&477.82\\
&&0.3772&0.3967&0.3287&0.3287&0.3182&---&974.75\\
\cline{2-9}

&\multirow{4}{*}{4}&0.3294&0.3524&0.2632&0.2544&0.2649&114&96.42\\
&&{0.3340}&{0.3599}&{0.2807}&{0.2632}&{0.2667}&91&100.06\\
&&0.3470&0.3763&0.2895&0.2749&0.2833&69&111.35\\
&&0.3026&0.3264&0.2193&0.2368&0.2377&---&  440.33\\
\cline{2-9}

&\multirow{4}{*}{5}&0.3397&0.3667&0.2568&0.2523&0.2662&74&80.84\\
&&{0.3422}&{0.3687}&{0.2703}&{0.2568}&{0.2676}&65&81.22\\
&&0.3525&0.3782&0.2703&0.2703&0.2743&44& 98.57\\
&&0.3295&0.3482&0.2568&0.2613&0.2527&---&300.75\\
\hline

\multirow{12}{2cm}{\texttt{MRF\_tpAll} \texttt{MRF\_tpS} \texttt{MRF\_oracle} \texttt{MRF\_tpNo}}

&\multirow{4}{*}{3} &0.4791&0.4883&0.4056&0.3916&0.3811&143&330.00\\
&&{0.4855}&{0.4924}&{0.4266}&{0.4103}&{0.3916}&113&346.53\\
&&0.5261&0.5362&0.4965&0.4592&0.4238&69&443.31\\
&&0.4558&0.4480&0.3706&0.3473&0.3608&---&903.79\\
\cline{2-9}

&\multirow{4}{*}{4} &0.4201&0.4365&0.4298&0.3333&0.3246&114&77.37\\
&&{0.4238}&{0.4416}&0.4123&{0.3450}&{0.3289}&97&77.92\\
&&0.4516&0.4869&0.4561&0.3919&0.3526&66&112.86\\
&&0.3397&0.3514&0.1930&0.2427&0.2693&---& 397.00\\
\cline{2-9}

&\multirow{4}{*}{5}&0.3760&0.3953&0.2535&0.2864&0.2958&71& 80.91\\
&&0.3810&{0.4003}&{0.2958}&{0.2911}&0.2887&68&81.01\\
&&0.4107&0.4395&0.3380&0.3521&0.3268&40&110.77\\
&&0.3338&0.3502&0.2394&0.2488&0.2549&---& 321.98\\
\hline

\multirow{12}{2.2cm}{\texttt{BM25TP\_tpAll} \texttt{BM25TP\_tpS} \texttt{BM25TP\_oracle} \texttt{BM25TP\_tpNo}}

&\multirow{4}{*}{3}&0.4128&0.4302&0.3310&0.3615&0.3493&142&352.71\\
&&0.4145&0.4354&0.3380&0.3662&0.3514&107&424.49\\
&&0.4300&0.4575&0.4085&0.4061&0.3683&67& 489.07\\
&&0.3807&0.4008&0.3380&0.3404&0.3275&---&912.59\\

\cline{2-9}

&\multirow{4}{*}{4}&0.3460&0.3542&0.2870&0.2580&0.2635&115&97.84\\
&&0.3506&0.3621&0.3130&0.2696&0.2730&94&150.02\\
&&0.3679&0.3931&0.3130&0.3014&0.2939&52&206.76\\
&&0.3052&0.3273&0.2261&0.2435&0.2426&---&410.79\\

\cline{2-9}

&\multirow{4}{*}{5}&0.3533&0.3580&0.2162&0.2703&0.2905&74&85.35\\
&&0.3712&0.3832&0.2703&0.2973&0.2986&57 &91.94\\
&&0.3894&0.4036&0.3378&0.3288&0.3149&30&140.26\\
&&0.3425&0.3587&0.2838&0.2928&0.2662&---&306.36\\

\hline

\end{tabular}
\end{table*}

% Becky: I'd suggest leaving this out; it's not that much better, and it's not always better.
% We also observe that \texttt{bm25\_b} performs slightly better than \texttt{bm25\_g}.

%There is an inverse correlation between the throughput and the number of position accessed. Table~\ref{ta:GOV_res} includes the time of labeling by the prediction models, which is no more than $0.01$ sec.\ in our experiments.  This would have less impact when the query set becomes larger.%

%The two machine learning methods, BP-ANN and GBDT, offer two different classification methods that both benefit scoring.  Table~\ref{ta:prec_rec} indicates BP-ANN results in better recall whereas GBDT results in better precision. Furthermore, BP-ANNs with different parameters provide many options.%, a trade-off between the performance and the effectiveness of query processing.

\section{Concluding remarks}\label{se:conc}

Recent studies have achieved promising retrieval performance by taking term proximity into consideration in relevance scoring. In this work, we propose a modified TP score ranking scheme which predicts which queries will benefit from using TP score in their rankings. In this way, we can: (a) achieve a better ranking from utilizing TP scores, and (b) achieve rankings with slightly better quality than the rankings given when always incorporating the TP score, but with better throughput.  In essence, we utilize TP score only when it's helpful.

Our work could be extended in several directions, e.g.: (a) The use of more features, particularly those that capture a notion of term proximity, could be explored. (b)  We could use a more complicated weighting of the queries' benefit from using TP score (here we use a simple $1$ vs.\ $0$ weighting).  This would enable us to use a linear regression model, which may achieve more effective results.  (c) Since different features benefit different types of queries, we could train a collection of models, individually designed for a single query type.

\section{ACKNOWLEDGMENTS}
This work is partially supported by NSF of China (grant numbers: 61373018, 11301288, 11550110491), Program for New Century Excellent Talents in University (grant number: NCET130301)  and the Fundamental Research Funds for the Central Universities (grant number: 65141021).

\bibliographystyle{abbrv}
\bibliography{sigproc}

\end{document}